**Direct Evidence for Two-Fluid Effects in Molecular Clouds**

By

David A. Tilley[1] (dtilley@nd.edu) and Dinshaw S. Balsara[1] (dbalsara@nd.edu)

**Running Head:** Direct Evidence for Two-Fluid Effects in Molecular Clouds

**Mailing Address:**

1: Department of Physics
225 Nieuwland Science Hall
University of Notre Dame
Notre Dame, Indiana 46556, USA

**Phone:** (574) 631-2712
**Fax:** (574) 631-5942

## Abstract

We present a combination of theoretical and simulation-based examinations of the role of two-fluid ambipolar drift on molecular line widths. The dissipation provided by ion-neutral interactions can produce a significant difference between the widths of neutral molecules and the widths of ionic species, comparable to the sound speed. We demonstrate that Alfvén waves and certain families of magnetosonic waves become strongly damped on scales comparable to the ambipolar diffusion scale. Using the RIEMANN code, we simulate two-fluid turbulence with ionization fractions ranging from $10^{-2}$ to $10^{-6}$. We show that the wave damping causes the power spectrum of the ion velocity to drop below that of the neutral velocity when measured on a relative basis. Following a set of motivational observations by Li & Houde (2008), we produce synthetic line width-size relations that shows a difference between the ion and neutral line widths, illustrating that two-fluid effects can have an observationally detectable role in modifying the MHD turbulence in the clouds.

## 1) Introduction

Most of the star formation in our Galaxy is thought to take place in molecular clouds, and a majority of it takes place in giant molecular clouds (GMCs). While GMCs are observed to be turbulent on the largest scales of parsecs, the theoretical models for how stars might be formed are still under debate since observations have not reached the point where they can help fully distinguish between competing models. One school of thought (see the reviews of Mac Low & Klessen 2004, and Ballesteros-Paredes et al. 2007, and the references therein) suggests that the turbulence provides pressure support to the GMC. Converging shocks induced in the turbulence can then provide sites for star formation, although the turbulence may also disrupt the cores that form. An alternative line of thought (Fiedler & Mouschovias 1993) asserts that cores collapse to form stars quasistatically with the physics of the collapse being dominated by microphysical processes associated with a two-fluid plasma (Tassis & Mouschovias 2005). An intermediate viewpoint consists of acknowledging the importance of turbulence while recognizing the dissipative effects that occur in a partially ionized plasma (Basu et al. 2009). Li & Houde (2008, LH hereafter) presented some very interesting line profile data that suggests that the important role of ambipolar diffusion in modifying the turbulence can indeed be distinguished observationally. The interpretation provided by LH was qualitative. If the results of LH could be supported by detailed dynamical calculations, it would indicate that two-fluid effects indeed play a very important role in modifying the turbulence on the smaller (inner) scales of the turbulence, thus setting the stage for core collapse on yet smaller scales. The goal of the present paper is to provide dynamical calculations that confirm LH's interpretation of their data.

Molecular clouds are primarily composed of neutral gas, but a small fraction of the material can remain ionized. This ionized component can couple to the magnetic field through the Lorentz force, and to the neutral cloud material via a frictional force. The intermediary role of the ions serves as a source of dissipation for turbulence, and can be characterized by an ambipolar dissipation scale $L_{AD} = v_A / \gamma \rho_i$ (where $v_A$ is the Alfvén speed in the coupled fluid, $\gamma$ is a frictional coupling coefficient between the neutral and ionized fluids, and $\rho_i$ is the density of the ions). This in turn allows one to define an ambipolar Reynolds number on the scale L, $R_{AD} = v_{RMS} L / v_A L_{AD}$ (Balsara 1996). Near the ambipolar dissipation scale, some MHD wave families may become damped or modified by the frictional coupling. There have been several attempts to analytically study the propagation of these waves in two-fluid magnetized plasmas that occur in molecular clouds (Langer 1978, Balsara 1996). In particular, the Alfvén waves are always strongly damped on scales smaller than the ambipolar dissipation scale, while either the fast magnetosonic or the slow magnetosonic waves will be damped below the dissipation scale, depending on whether the fluid is pressure or magnetically dominated. Measurements of the strengths of magnetic fields via the Zeeman effect or the Chandrasekhar-Fermi method suggest that the magnetic pressure is typically comparable to the total gas pressure (Crutcher 1999, 2004; Jijina et al. 1999). As a result, in a highly turbulent cloud pressure and magnetic field fluctuations will lead to some regions that are pressure-dominated and others that are magnetically dominated.

LH examined the possibility that the existence of the ambipolar dissipation scale might be inferred from the intercomparison of optically thin neutral and ion line profiles. They measured the line widths of $HCO^+$ , a tracer of the ionized component, and HCN, a tracer of the neutral component, at different spatial resolutions in the massive star-forming region M17. These two species have similar molecular weights and similar critical densities for emission, simplifying their comparison. LH reported that the minimum line widths of $HCO^+$ observed at any scale were systematically smaller than the line widths of HCN by approximately 0.5 km $s^{-1}$. They further conjectured that the minimum line widths on any given scale is related to a power law that connects the turbulent line widths to the ambipolar diffusion Reynolds number.

On the largest scales in any turbulent plasma, one should expect turbulent driving to produce equipartition between the ionized and neutral components. Consequently, line profiles of ions and neutrals measured on these scales should have the same velocity broadening. On the smaller scales, where ambipolar diffusion becomes significant, the dispersion analysis for wave propagation (Balsara 1996) says that certain families of waves in the ionized fluid should be damped. As a result, the RMS velocity in the neutral fluid should be larger than the RMS velocity in the ionized fluid on smaller length scales. LH relied on this fact in order to understand their data within the context of a turbulence model. To understand those observations, one needs to carry out a driven two-fluid turbulence simulation, and examine the RMS velocities in the ions and neutrals on a range of length scales. Ideally, one would like to have a large separation of scales between the turbulent driving and the dissipation. It is not currently feasible to carry out such computations with a large dynamic range and a realistic ionization fraction, however two-fluid calculations with some dynamic range and an ability to fully capture the dissipation scales are now possible. We present such calculations and verify that the trends noticed by LH are indeed seen in the simulations. A more detailed examination of the line widths on a range of scales is, therefore, carried out showing that the systematics observed by LH are in fact verified. As a result, we have a direct probe of the ambipolar dissipation scales which would otherwise be unobservable.

Numerical methods than can correctly capture the dissipation are essential in order to track all of the different wave families on scales larger and smaller than the dissipation scale. The numerical investigation of partially ionized magnetized fluids has proceeded primarily on two algorithmic paths:

1) A two-fluid treatment, which evolves the ionized and neutral components separately and utilizes a friction term to couple them. The large Alfvén velocities are usually compensated for by invoking a "heavy ion approximation (HIA)" (Oishi & Mac Low 2006, Li et al. 2008), which inflates the ion masses in order to reduce the Alfvén velocity while simultaneously reducing the coupling coefficient to keep the friction term unchanged. Tilley & Balsara (2008) have calibrated the two-fluid treatment when the HIA is dispensed with and have shown that it is possible to saliently simulate plasmas with ionization fractions of $10^{-6}$ using their RIEMANN code.

2) A single-fluid treatment, in which the advection terms are removed from the ion momentum equation in order to express the ion-neutral friction force in terms of the magnetic stresses. The resulting equations closely resemble the ideal MHD equations, but with an additional diffusive term in the induction equation; see O'Sullivan & Downes (2006, 2007) for numerical work along this direction. In this approach one cannot extract density fluctuations for the ions. Furthermore, the plasma effects below the ambipolar diffusion scale are obliterated by the numerical approximation (Balsara 1996). As a result, this latter approach is not suited for extracting line profiles in the ions.

Both the single-fluid approach and the HIA for the two-fluid method either strongly modify or ignore the ion momentum equation, preventing either of these strategies from producing separate diagnostics of the ions and neutrals. The two-fluid method without the HIA can produce this information if one is willing to absorb the cost of a restrictive time step that results from the Alfvén and magnetosonic waves. Furthermore, the HIA and the single-fluid treatment will both modify the propagation and dissipation characteristics of the MHD waves (Balsara 1996).

The theoretical background used in the analysis of LH relied on the well-ionized, single-fluid calculations of Ostriker et al. (2001). A similar set of simulations incorporating two-fluid turbulence was performed by Oishi & Mac Low (2006), using a relatively large ionization fraction of 0.1. Oishi & Mac Low (2006) and Li et al. (2008) discovered that turbulent structures could still be maintained at scales smaller than the ambipolar diffusion scale via the propagation of slow magnetosonic waves. This would appear to undermine the proposed explanation of Li & Houde (2008). However, the dispersion analysis of Balsara (1996) and the results from Tilley & Balsara (2008) show that as the ionization fraction becomes much smaller, there is a large range of wave numbers where the only propagating waves are sound waves in the neutral fluid. One might expect that if two-fluid turbulence calculation were performed at ionization fractions of $10^{-4}$ or less, a more pronounced effect due to ambipolar diffusion might be observed.

In this Letter, we present a series of isothermal two-fluid MHD simulations with turbulent forcing, performed at ionization fractions ranging from $10^{-2}$ to $10^{-6}$. We demonstrate that there is a pronounced difference in the line widths of the ionized and neutral fluids that is enhanced at smaller ionization fractions. Section 2 presents our methods, Section 3 presents the results and Section 4 presents our conclusions.

## 2) Methods

We use the RIEMANN code (Balsara 1998a,b; Balsara & Spicer 1999a,b; Balsara 2004) to update the hydrodynamic and MHD fluid variables. We incorporate ion-neutral friction via an operator-split method (Tilley & Balsara 2008).

Our initial velocity distribution is generated in Fourier space, with an initial spectrum of $\exp(-k^2)\, k^{-5/3}$ and complex amplitudes drawn from a Gaussian random distribution. We chose our initial conditions such that the dissipation scale is found within the computational domain for ionization fraction of $10^{-4}$, as the computational expense increases dramatically at lower ionization fractions. In the context of dense molecular cloud cores, the physical scale of the ambipolar dissipation scale is extremely small at this ionization fraction, on the order of μpc. This is much smaller than the scales observed by LH, who assumed a much smaller ionization fraction of ~$10^{-7}$ in their estimations of the dissipation scale. Those densities are computationally unfeasible at present. We are free, however, to rescale our simulation to mean total densities of $10^{-3}$, with a corresponding increasing of the ambipolar dissipation scale to mpc ranges. We describe our initial neutral densities, molecular weights, sound speeds and ambipolar diffusion drag coefficients for both density scalings in Table 1. The initial ionization fractions are in Table 2, along with the ambipolar dissipation scale for each simulation. The initial magnetic field is set so that the Alfven velocity of the neutral+ionized fluid is $\mathrm{v}_A = (1+\rho_i/\rho_n)^{-1/2} c_s$ , and is uniform in the x-direction.

We provide forcing with a solenoidal vector field to the kinetic energy of the neutral fluid with a tapered spectrum centered on $kL/2\pi = 4$. For the simulations at ionization fractions of $10^{-4}$ and larger, the forcing scale is larger than the dissipation scale and we have a clear separation between the forcing and dissipation scales. For ionization fractions of $10^{-5}$ and $10^{-6}$, it is not possible to have this requisite scale separation. These simulations were performed on meshes with $192^3$ zones. Because the problem demands a very large dynamic range, we do not make hard claims about spectra, but only extract the information that robustly permeates all of our simulations. We will show that this is sufficient to reproduce the results of LH on scales near the dissipation scale. The results presented here are taken after the turbulence has reached steady state at one turnover time.

## 3) Results

This section is divided into three parts. We first discuss the dispersion relation for a two-fluid plasma and show its consequences for the turbulent spectra. Next we show simulated line profiles for the ions and neutrals. Finally, we show that our simulations reproduce the results of LH.

### 3.1) Dispersion Analysis and Spectra

Balsara (1996) and Tilley & Balsara (2008) studied the dispersion analysis for a two-fluid, self-gravitating, isothermal system. They demonstrated that below the ambipolar dissipation scale, only one family of magnetosonic waves remain undamped (the fast magnetosonic waves persist if the Alfvén speed is less than the sound speed of the fluid; the slow magnetosonic waves persist if the Alfvén speed is greater than the sound speed). The other magnetosonic wave and the Alfvén wave are strongly damped. Balsara (1996) also demonstrated that there is a smaller scale on which the neutral and ionized fluids are sufficiently decoupled that the Alfvén and magnetosonic waves can re-emerge in the ionized fluid only. In this weakly coupled regime, the Alfvén and magnetosonic velocities depend only on the ion density; this is is contrast to the strongly-coupled regime on scales larger than the dissipation scale, where the relevant velocities depend on the combined density of the two fluids. The effects of the HIA on the propagation of MHD two-fluid waves has not been studied, so we briefly present that here.

We draw on the analysis of Balsara (1996), which presents the eigenvalue equation for the two-fluid isothermal system. We will compare the dispersion relation for a gas with an ionization fraction of $10^{-6}$ with and without the HIA. We increase the molecular weight of the ions by factor of $10^{4}$, corresponding to the HIA. The HIA correspondingly decreases the coupling coefficient by the same factor of $10^{4}$. Since the ambipolar dissipation scales varies with the product of the ion density and the coupling coefficient, it remains unchanged under the HIA. However, the ion momentum equation is modified due to the increased ion inertia, and this modification leads to a change in the propagation of the MHD waves. In Figure 1a, we show the phase velocity of MHD waves on scales above and below the dissipation scale for an ionization fraction of $10^{-6}$ without the HIA, for a wave traveling at an angle of 31° with respect to the mean magnetic field which has a strength set by $v_A = 1.5\ c_s$. We plot the slow magnetosonic waves that turn into sound waves with solid lines, the slow magnetosonic waves that appear in the ions only with dotted lines, the Alfvén waves with dashed lines, and the fast waves with dot-dashed lines. Due to the large range wave speeds present, we split the plot into three sections that show the three key ranges in speed that are of interest – the range [-2 $c_s$, 2$c_s$] is in the center, [300 $c_s$, 450 $c_s$] above and [-450 $c_s$, -300 $c_s$] on the bottom. We clearly see that the fast and Alfvén waves disappear on the dissipation scale $kL_{AD} \sim 1$. An examination of the damping rate (Balsara 1996) shows that these waves propagate with strong damping on length scales that are almost an order of magnitude larger than $L_{AD}$. As a result, it is not surprising that LH detect a velocity difference between the ions and neutrals on scales that are an order of magnitude larger than their conjectured dissipation scales. At $kL_{AD} \sim 100$, we see the re-emergence of these two waves, now unhindered by the neutral inertia. In contrast, in Figure 1b we show the results for the HIA. Here, there is no dissipation range to be found in between where the MHD waves die off and where they re-appear in the weakly coupled regime. The dissipation scale in Figure 1b is the correct value for an ionization fraction of $10^{-6}$, but the pattern of the dispersion relation in fact looks very much like the dispersion relation for a fluid at an ionization fraction of $10^{-2}$. The HIA in this case will not correctly capture the MHD waves within the dissipation range.

Figure 2 shows the turbulent spectrum for the kinetic energy of the neutrals (dotted line), kinetic energy of the ions (solid line), and magnetic field (dot-dashed line). Figure 2a shows this data for our simulation with an ionization fraction of $10^{-4}$, and Figure 2b shows the same for an ionization fraction of $10^{-5}$. The spectra have been normalized so that they have the same value at the largest scale in the simulation. We see from Figure 2a that on scales comparable to the dissipation scale (i.e. on length scales that have $R_{AD}$~1), the kinetic energy in the ions is smaller than the kinetic energy in the neutrals on a relative basis. On the smaller scales of any high-Mach-number MHD calculation, much of the energy is expected to be in the form of waves, not shocks. Since our dispersion analysis shows that the Alfvén waves as well as one family of magnetosonic waves is strongly damped, the magnetic energy also shows a rapid decrement below the dissipation scale. Figure 2b, which pertains to an ionization fraction of $10^{-5}$, but is similar to Figure 2a in all other respects, shows a larger range of scales where the kinetic energy of the ions as well as the magnetic energy is damped relative to the kinetic energy of the neutrals. Because of resolution limitations, we had to force this simulation on scales that are smaller than the dissipation scale. Had it been forced on larger scales, the spread between the ion and neutral kinetic energies would have been even more pronounced. The simulation with ionization fraction of $10^{-2}$, by contrast, has a very small $L_{AD}$. Consequently, the velocity spectra for ions and neutrals closely track each other. This is consistent our wave propagation model from Figure 1a and our prior demonstration that the dissipation scale has a larger value at smaller ionization fractions. On going to even smaller ionization fractions, this trend becomes even more pronounced.

### 3.2) Simulated Line Profiles

The shapes of spectral line profiles provide one of the few methods to probe the kinematics within a cloud because they directly track the velocities in the turbulent fluid. In Figure 3 we plot simulated line profiles for our runs, integrated over the entire domain. We calculate the profiles in Figure 3 by summing individual lines from every cell with an emissivity proportional to the density, a line width equal to the isothermal sound speed, and centered on the mean velocity parallel to the line-of-sight within that zone. Figure 3a and 3b show the line profiles for our simulations with ionization fractions of $10^{-4}$ and $10^{-5}$. We see that the equivalent width of the neutral line profile (marked by a dashed line) is wider than the equivalent width of the ionized line profile (solid line), by 1.2 $c_s$ for Figure 3a and 1.9 $c_s$ for Figure 3b. For ionization fractions of $10^{-2}$, $10^{-3}$ and $10^{-6}$ we find that the equivalent widths of the neutral line profiles are larger than that of the ions by 0.016 $c_s$, 0.33 $c_s$, and 2.1 $c_s$ respectively. The claim in LH is that they are looking at the ambipolar dissipation scales for two-fluid turbulence, and our simulations focus on the dissipation scales. Our spectra have shown that on those scales we should expect a smaller RMS velocity in the ions and this is supported by the dynamics of wave propagation. The simulated line profiles that we get are also consistent with that expectation. We thus find that the results reported by LH of a systematic shift between the widths of ionized and neutral lines could be justified by our simulations.

### 3.3 Reproducing the Results of LH via Simulations

In order to test the hypothesis of LH, we calculate individual line profiles along each column of zones in our simulations using the same prescription as we used in Section 3.2. We then combined the projected line profiles by binning adjacent profiles together, in order to get a distribution of line profiles on different scales. In analogy to LH, we then plot the square of the line widths as a function of scale for ionization fractions of $10^{-4}$ and $10^{-5}$ in Figures 4a and 4b, respectively. The ionized component line widths are marked by squares, and the neutral component line widths are marked by diamonds. As the ion and neutral line profiles are measured on exactly the same scales as each other, we introduce a slight offset to the neutral line widths in the x-axis in order for the differences in the distributions of the two sets of data to become apparent. We have a larger range of scales to work with as well, so the x-axis is logarithmic in Figure 4.

LH claim that the lower envelope of line widths for the neutral fluid is systematically ~0.5 $km^2\ s^{-2}$ larger than the lower envelope of line widths for the ionized fluid. We see that the lower envelopes differ typically by ~ 1-2 $c_s^2$ for ionization fractions of $10^{-4}$, and up to 3 $c_s^2$ on the largest scales at ionization fractions of $10^{-5}$. The sound speed of the gas at a temperature of 50K is 0.42 km $s^{-1}$; the differences in the lower envelope of the line widths thus works out to about 0.54 $km^2\ s^{-2}$. The difference in the line widths is more pronounced at lower ionization fractions. Our simulations produce a result very similar to LH, thus confirming that two-fluid turbulence can explain the observational results.

## 4) Conclusions

The large observed line widths in the ions and neutrals are most conveniently attributed to MHD turbulence in molecular clouds. Since the plasma in a molecular cloud is partially ionized, it will display two-fluid effects. The low levels of ionization in the plasma ensure that the ionized and neutral fluids become partially uncoupled on sub-parsec scales. This process causes certain families of magnetosonic waves and all families of Alfvén waves to be strongly damped on scales comparable to the ambipolar diffusion scale, $L_{AD}$. As a result, the power spectrum in the ions dips below that of the neutrals when measured on a relative basis. Consequently, we show that line profiles for optically thin neutral lines are wider than the line profiles for optically thin ion lines. The above statement holds true as long as the ion and neutral species have comparable molecular weights. We use this to produce synthetic line width-size relations that reproduce the trends in LH's observed data. The observations of LH, taken along with our simulations, therefore provide direct evidence that two-fluid effects play an important role in modifying the MHD turbulence in molecular clouds. This modification is scientifically significant because it takes place on the same scale that cores are formed. As a result, models for star formation should include two-fluid effects without resorting to computationally expedient simplifications.

The ALMA instrument will be an even more powerful and precise tool for measuring line widths as well as magnetic fields. As a result, future observational studies should be able to demonstrate the importance of two-fluid effects in several systems using the methods pioneered by LH and computationally supported by this paper. Future observational studies should be able to study ion and neutral lines at a range of densities and temperatures (Bergin & Tafalla 2007). This makes a very good case for including chemical networks in computational models of molecular cloud turbulence.

DSB acknowledges support via NSF grants AST-0607731 and NSF-AST-0947765, and NASA grants NASA-NNX07AG93G and NASA-NNX08AG69G. The authors are also grateful for support from the CRC at UND.

Table 1: Initial conditions for the simulations.

| Neutral density ($cm^{-3}$) | Sound speed ($km\ s^{-1}$) | Drag coefficient ($cm^3\ g^{-1}\ s^{-1}$) | Ion molecular weight | Neutral molecular weight | $v_{rms}$ | $L_{box}$ |
|---|---|---|---|---|---|---|
| $10^6$ | 0.424 | $3.5x10^{13}$ | 29.0 | 2.3 | 3 $c_s$ | 0.31 μpc |
| $10^3$ | 0.424 | 3.5x1013 | 29.0 | 2.3 | 3 $c_s$ | 0.31 mpc |

Table 2: List of simulations.

| Run | R2 | R3 | R4 | R5 | R6 |
|---|---|---|---|---|---|
| $\xi$ | $10^{-2}$ | $10^{-3}$ | $10^{-4}$ | $10^{-5}$ | $10^{-6}$ |
| $L_{AD}/L_{box}$ | $2.64x10^{-3}$ | $2.64x10^{-2}$ | 0.264 | 2.64 | 26.4 |
| $R_{AD}$ ($L_{box}$) | $1.13x10^3$ | $1.13x10^2$ | 11.3 | 1.13 | 0.113 |

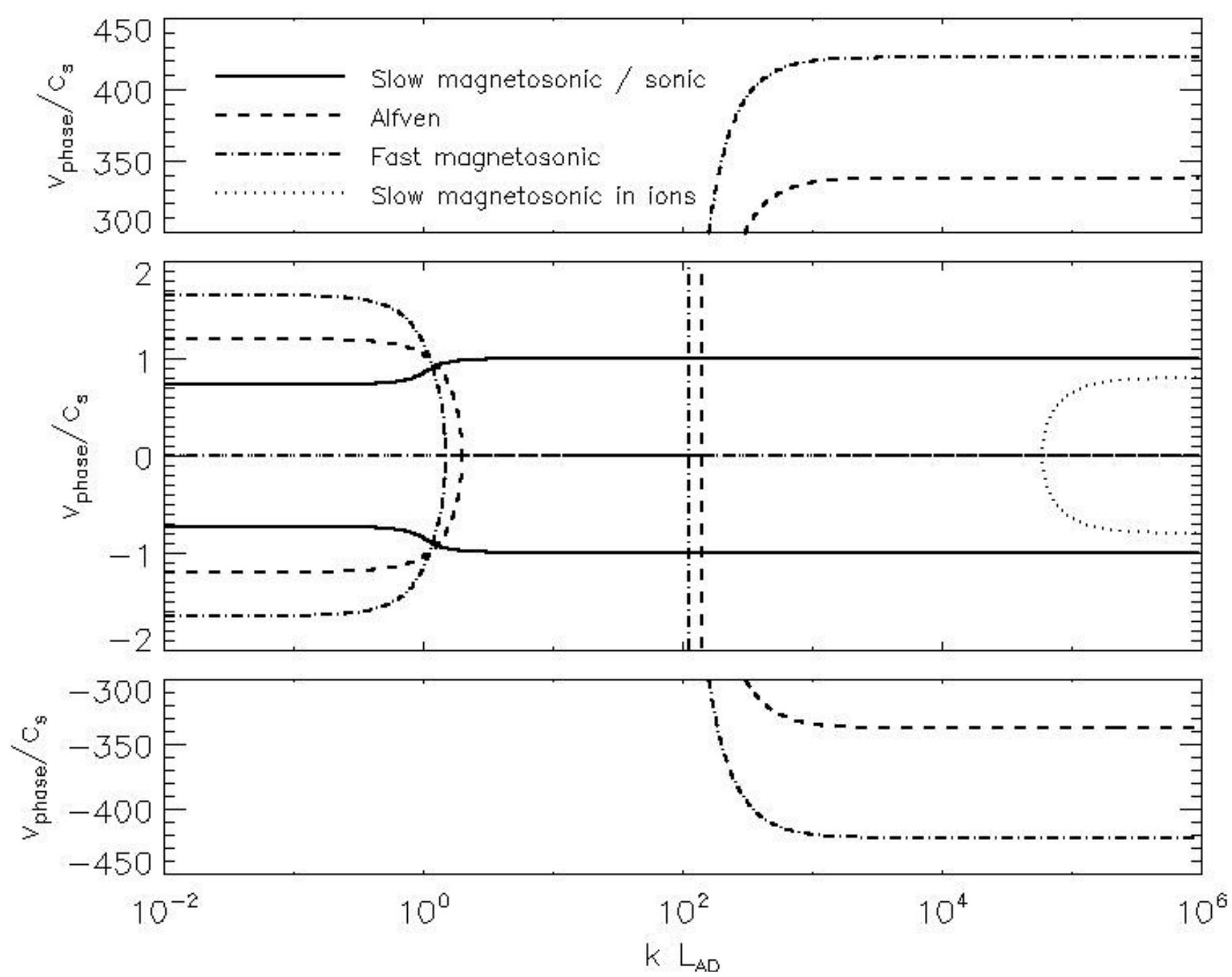

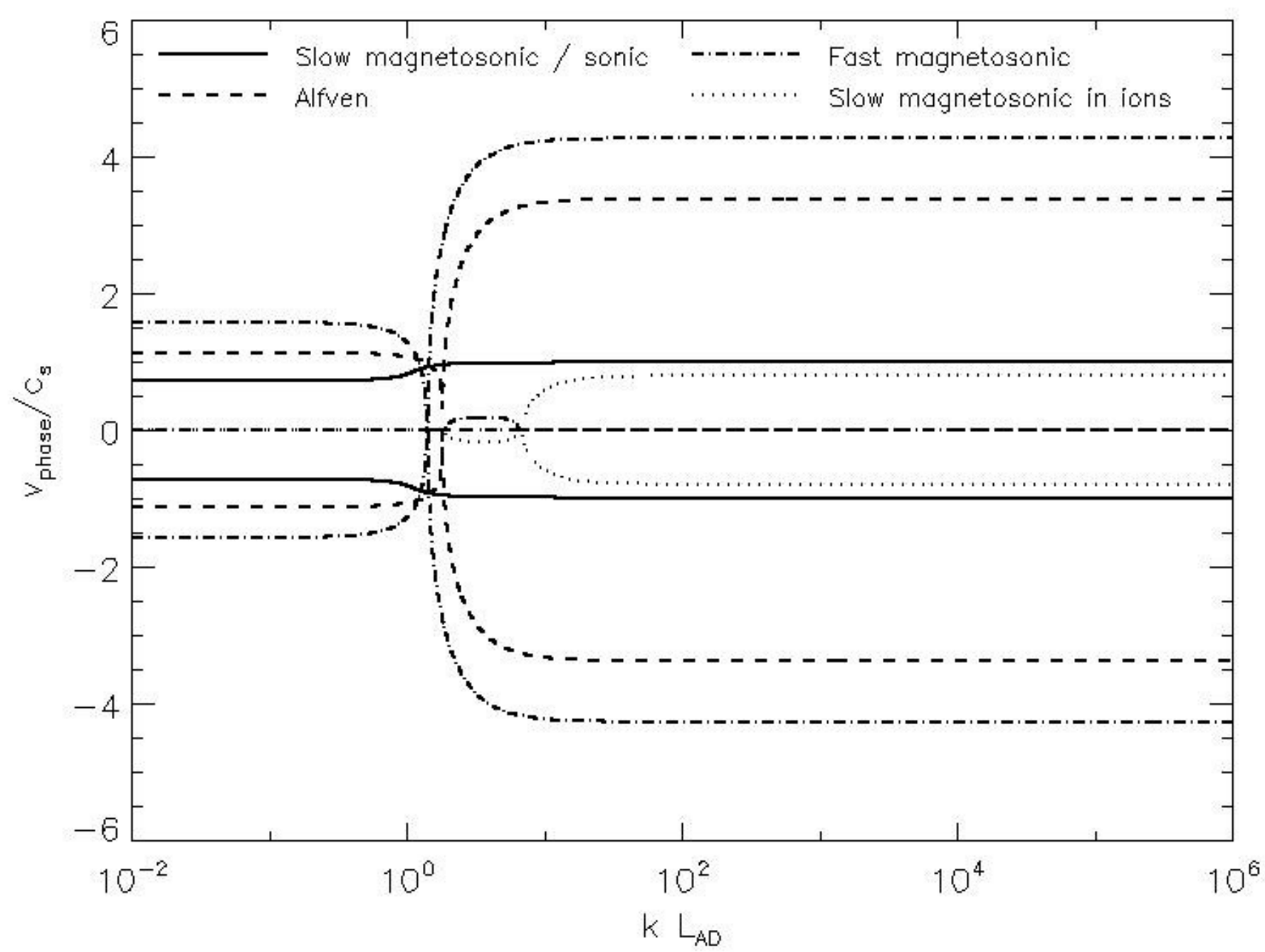

Figure 1: (a) Real part of the dispersion relation for an ionization fraction of $10^{-6}$. In the weakly-coupled regime ($kL_{AD}$ > 100), the Alfvén speed depends solely on the ion density, and hence is significantly larger than the Alfvén speed in the strongly-coupled regime. As a result, we show these in the upper and lower panels.

(b) Real part of the dispersion relation for the heavy-ion approximation.

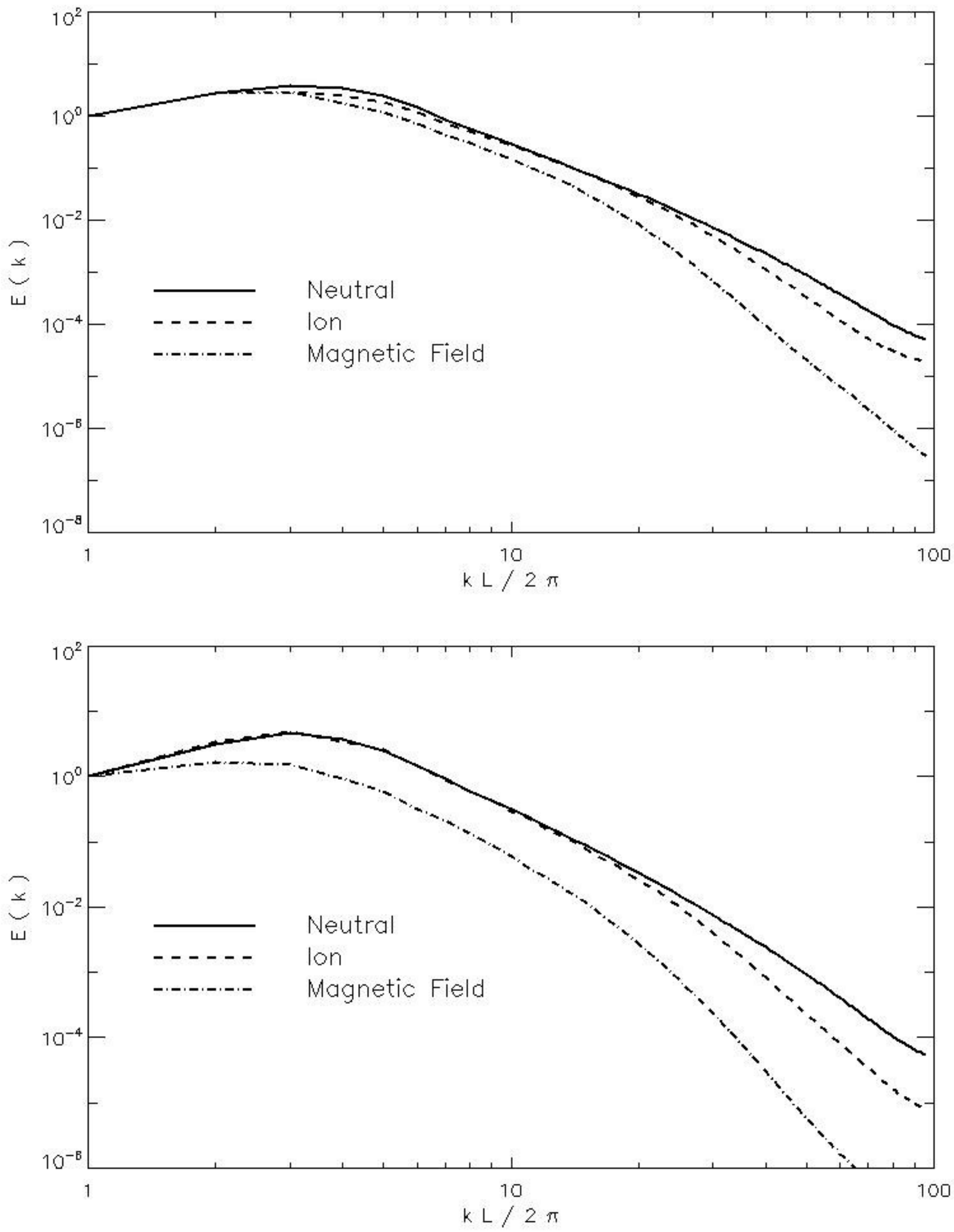

Figure 2: (a) Spectrum of the simulation with an ionization fraction of $10^{-4}$. (b) Spectrum of the simulation with an ionization fraction of $10^{-5}$.

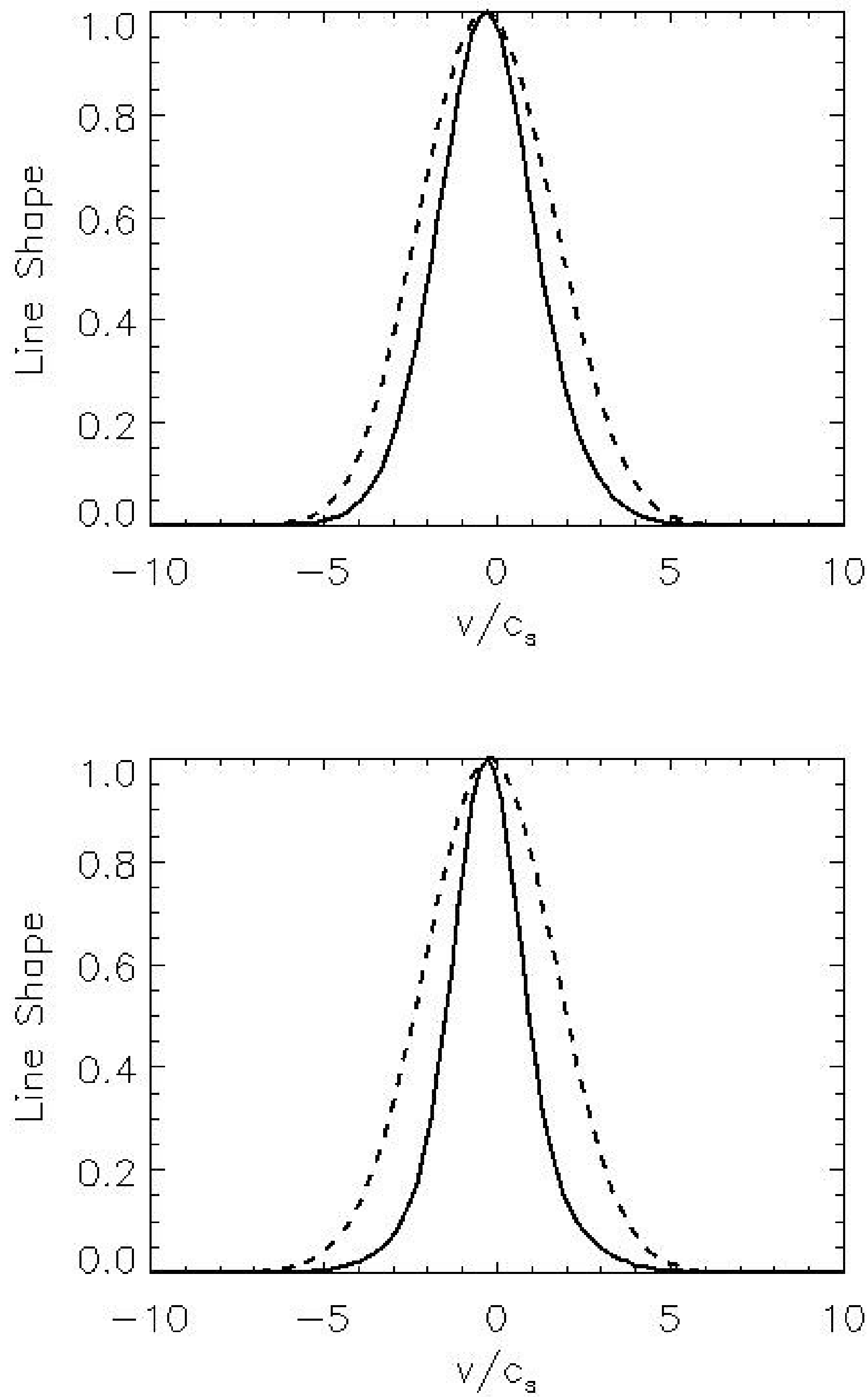

Figure 3: (a) Line profile of the ionized fluid (solid line) and neutral fluid (dashed line), integrated over the entire volume and normalized to the same height, for an ionization fraction of $10^{-4}$. (b) Line profile for an ionization fraction of $10^{-5}$.

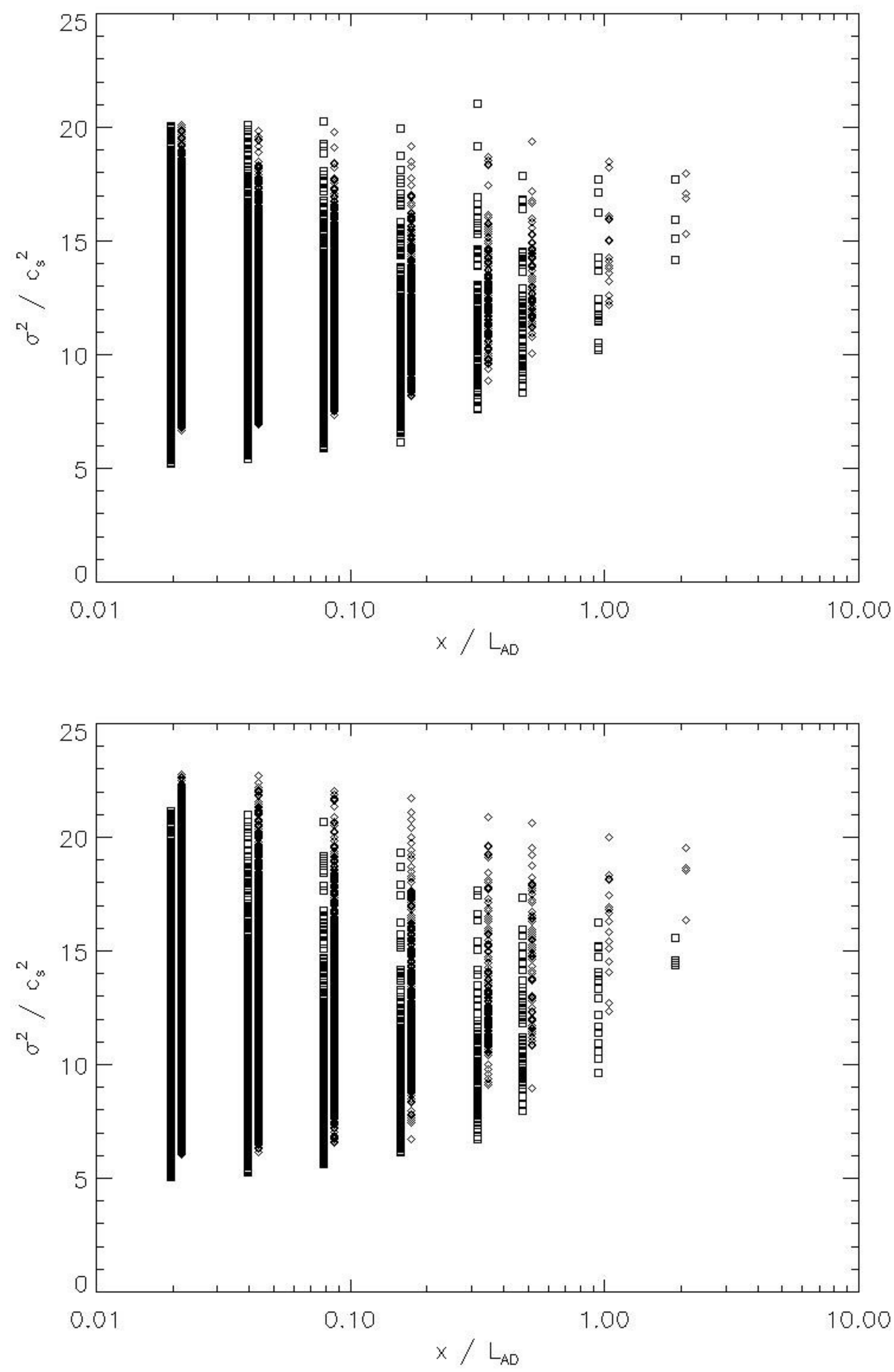

Figure 4: (a) Line width-size relation for the simulation with an ionization fraction of $10^{-4}$. The ionized fluid is marked by squares. The neutral fluid is marked by diamonds, and is offset by a small amount in order to show the differences in the line widths of the neutral and ionized fluids on the same scale. (b) Line width-size relation for the simulation with an ionization fraction of $10^{-5}$.